\documentclass[12pt]{article}
\usepackage{epsfig}
\usepackage{latexsym}
\usepackage{color}
\usepackage{nicefrac}
\usepackage[latin1]{inputenc}

\newcommand{\mysquare}[0]{\raise-.2ex\hbox{{\Large$\Box$}}}
\def\lsim{\mathrel{\rlap {\raise.5ex\hbox{$ < $}}
{\lower.5ex\hbox{$\sim$}}}}
\def\gsim{\mathrel{\rlap {\raise.5ex\hbox{$ > $}}
{\lower.5ex\hbox{$\sim$}}}} \topmargin -1.5cm \textheight=22.5cm
\catcode`\@=11
\newcount\hour
\newcount\minute
\newtoks\amorpm
\hour=\time\divide\hour by60
\minute=\time{\multiply\hour by60 \global\advance\minute by-\hour}
\edef\standardtime{{\ifnum\hour<12 \global\amorpm={am}%
        \else\global\amorpm={pm}\advance\hour by-12 \fi
        \ifnum\hour=0 \hour=12 \fi
        \number\hour:\ifnum\minute<10 0\fi\number\minute\the\amorpm}}
\edef\militarytime{\number\hour:\ifnum\minute<10 0\fi\number\minute}
\def\draftlabel#1{{\@bsphack\if@filesw {\let\thepage\relax
   \xdef\@gtempa{\write\@auxout{\string
      \newlabel{#1}{{\@currentlabel}{\thepage}}}}}\@gtempa
   \if@nobreak \ifvmode\nobreak\fi\fi\fi\@esphack}
        \gdef\@eqnlabel{#1}}
\def\@eqnlabel{}
\def\@vacuum{}
\def\draftmarginnote#1{\marginpar{\raggedright\scriptsize\tt#1}}
\def\draft{\oddsidemargin -.2truein
        \def\@oddfoot{\sl preliminary draft \hfil
        \rm\thepage\hfil\sl\today\quad\militarytime}
        \let\@evenfoot\@oddfoot \overfullrule 3pt
        \let\label=\draftlabel
        \let\marginnote=\draftmarginnote
   \def\@eqnnum{(\theequation)\rlap{\kern\marginparsep\tt\@eqnlabel}%
\global\let\@eqnlabel\@vacuum}  }
\relax
\newcommand{\ba}[0]{\begin{eqnarray}}
\newcommand{\ea}[0]{\end{eqnarray}}
\def\bs{\begin{subequations}}
\def\es{\end{subequations}}

\def\thebibliography#1{%
\vskip 0.5cm \centerline{\bf References}
\list{%
[\arabic{enumi}]}{\settowidth\labelwidth{[#1]}
\leftmargin\labelwidth
\advance\leftmargin\labelsep
\usecounter{enumi}}
\def\newblock{\hskip .11em plus .33em minus .07em}
\sloppy\clubpenalty4000\widowpenalty4000
\sfcode`\.=1000\relax}

\renewcommand{\theequation}{\arabic{section}.\arabic{equation}}

\renewcommand{\section}{\setcounter{equation}{0}\@startsection%
{section}{1}{0mm}{-\baselineskip}{0.5\baselineskip}%
{\normalfont\normalsize\bfseries}}

\renewcommand{\subsection}{\@startsection%
{subsection}{2}{0mm}{-\baselineskip}{0.5\baselineskip}%
{\normalfont\normalsize\slshape}}

\usepackage{amscd} \usepackage{amsmath} \usepackage{amsfonts}
\def\thefootnote{\fnsymbol{footnote}}
\def\es{\end{subequations}}

\newcommand{\uarrw}[0]{\mathrel{
{\raise.5ex\vbox{\hrule width 1cm}\hskip-6pt\rightarrow}}}
\begin{document}
%\verb|\usepackage{draftcopy}|\\
\renewcommand{\theequation}{\arabic{section}.\arabic{equation}}
\begin{titlepage}
\begin{flushright}
CPTH-PC086.1106\\
hep-th/0612243\\
\end{flushright}
\begin{centering}
\vspace{45pt}
{\bf PARAFERMIONS, BRANE DISTRIBUTIONS AND FRW HIERARCHIES} $^\ast$\\
\vspace{30pt} {P.M. Petropoulos}\\
\vspace{15pt} {\it Centre de Physique Th{\'e}orique, Ecole
Polytechnique $^\dagger
$}\\
{\it 91128 Palaiseau Cedex, FRANCE}
 \vspace{22pt}

{\bf Abstract}\\
\end{centering}
\vspace{10pt}

\noindent I review a class of exact string backgrounds, which
appear in hierarchies, where the boundary of the target space of
an exact sigma model is itself the target space of another exact
model. From the worldsheet viewpoint this is due to the existence
of $(1,1)$ operators based on parafermions. From the target space
side, it is reminiscent of the structure of maximally symmetric
Friedmann--Robertson--Walker cosmological solutions, with broken
homogeneity though. Cosmological evolution in this framework
raises again the question of the nature of time in string theory.

\vfill
\begin{flushleft}
CPTH-PC086.1106\\
December 2006
\end{flushleft}
\hrule width 6.7cm \vskip.1mm{\small \small \small $^\ast$\ Based
on works with K. Sfetsos. To appear in the proceedings of the
mid-term meeting of the RTN \textsl{Constituents, fundamental
forces and symmetries of the universe (ForcesUniverse)}, Naples,
Italy, October 9 -- 13 2006. Research partially supported by the
EU under the contracts MEXT-CT-2003-509661, MRTN-CT-2004-005104
and MRTN-CT-2004-503369 and by the Agence Nationale pour la Recherche.\\
$^\dagger$\ Unit{\'e} mixte  du CNRS et de  l'Ecole Polytechnique,
UMR 7644.}
\end{titlepage}
\newpage
\setcounter{footnote}{0}
\renewcommand{\thefootnote}{\arabic{footnote}}

\setcounter{section}{0}
\section{Parafermions and brane deformations}\label{sec:par}

The investigation of string moduli space has been an important
issue for a variety of reasons. Exact worldsheet theories are good
starting points for further deformations, using integrable
marginal operators, whenever available. Conformal models based on
group manifolds $G$ and stabilized with the help of Kalb--Ramond
fields (Wess--Zumino--Witten models) offer the ideal arena for
investigating deformations. They possess a high degree of symmetry
($G\times G$) realized in terms of holomorphic and antiholomorphic
currents. Bilinears of the latter are marginal operators which
have been discussed prolifically in the literature
\cite{Forste:1994wp,Forste:2003km,Israel:2003ry,Israel:2003cx,
Israel:2004vv,Israel:2004cd,Petropoulos:2004ir,Israel}.

Gauged WZW models, hereafter called non-Abelian conformal cosets
$G/H$, are also exact conformal theories, with poor symmetry
though. The subgroup $H$ acts vectorially (or axially) and spoils
completely the symmetry, contrary to what happens in ordinary
geometric cosets, where the action is one-sided. The $(1,0)$ and
$(0,1)$ currents of the original WZW model do not survive but new
holomorphic and antiholomorphic operators appear, with remarkable
conformal and braiding properties: the \emph{parafermions}
\cite{paraf}. For compact groups, the anomalous dimensions of
these operators obey $h, \bar{h} < 1$.

It has been recently realized that \emph{the parafermions can be
used as building blocks for new marginal operators}, in a
framework which does also enable to clarify the brane
interpretation of the deformed background \cite{PS05}. The
starting point in this analysis is the gauged WZW model
${SL(2,\mathbb{R})}/{U(1)} \times {SU(2)}/{U(1)}$, which appears
as the non-trivial part of the target space of a continuous
distribution of NS5-branes on a circle, in the near-horizon limit
\cite{sfet1}. Deforming the circle into an ellipsis, preserves the
original supersymmetry and breaks half of the isometry
($SO(2)\times SO(2) \to SO(2)$). This deformation keeps the exact
nature of the solution and is driven by an identifiable marginal
operator which is a bilinear in the holomorphic and
antiholomorphic parafermions of ${SU(2)}/{U(1)}$ dressed with a
non-holomorphic conformal operator of the non-compact
${SL(2,\mathbb{R})}/{U(1)}$.

This result opened the way to a new class of integrable marginal
deformations triggered by non-left-right-factorizable operators,
available in sigma models without symmetries, namely gauged WZW
models. The class of models based on (pseudo)orthogonal groups was
studied in \cite{PS06}. Although it remains still unclear whether
a brane distribution exists, which could reproduce these
backgrounds as near-horizon geometries, these models are of
special interest because of their \emph{potential
%holographic and
cosmological applications}.

\section{FRW hierarchies in string theory}\label{sec:FRW}

In cosmology, isotropy and homogeneity of space imply the
existence of a co-moving frame where
\begin{equation}\label{eq:met}
  \mathrm{d}s^2 = - \mathrm{d}t^2 +a^2(t)\,{\gamma_{ij}(\mathbf{x})
\, \mathrm{d}x^i\, \mathrm{d}x^j}.
\end{equation}
The Euclidean metric $\gamma_{ij}$ is homogeneous and maximally
symmetric and describes therefore a geometric coset of a
(pseudo)orthogonal group. It solves three-dimensional Einstein's
equations with pure cosmological constant $\Lambda = R^{(3)}/6$,
where $R^{(3)}$ is the constant three-dimensional curvature
scalar. Three situations appear:
\begin{eqnarray}
S^3 : & {SO(4)}/{SO(3)} & \Lambda>0, \nonumber \\
H_3 : & {SO(3,1)}/{SO(3)} & \Lambda<0, \nonumber \\
E_3 : & \mathrm{flat \ space} & \Lambda=0. \nonumber
\end{eqnarray}
The scale factor $a(t)$ satisfies Friedmann--Lemaître equations.
In the absence of matter (pure cosmological constant) these
equations are again exactly solvable and the four-dimensional
space--time is also maximally symmetric with constant scalar
curvature $R^{(4)} = 4 \Lambda = 2 R^{(3)}/3$:
\begin{eqnarray}\nonumber
&\mathrm{dS}_4:  {SO(4,1)}/{SO(3,1)}&  \Lambda>0 \mathrm{\ with \ spatial \ sections\ } S^3  , \nonumber \\
&\mathrm{Einstein-dS}_4  & \Lambda>0 \mathrm{\ with \ spatial \ sections\ } E_3  , \nonumber \\
&\mathrm{AdS}_4: {SO(3,2)}/{SO(3,1)} & \Lambda<0 \mathrm{\ with \
spatial \ sections\ } H_3 . \nonumber
\end{eqnarray}
In all previous expressions, $S,H,\mathrm{dS}$ and $\mathrm{AdS}$
stand for spheres, hyperbolic planes, de-Sitter and anti-de Sitter
spaces.

This hierarchical structure in which four-dimensional maximally
symmetric space--times are foliated with three-dimensional
maximally symmetric spaces is intimately related to the underlying
orthogonal-group geometric-coset structure or equivalently to the
maximal symmetry.

Geometric cosets are not exact string\footnote{The
three-dimensional anti de Sitter and the three-sphere are
exceptions because they are also group manifolds.} backgrounds.
Supplemented with antisymmetric tensors, they solve the
supergravity equations -- with vanishing dilaton since they have
constant curvature -- and appear as factors of near-horizon
geometries of branes. As I mentioned in Sec. \ref{sec:par},
however, gauged WZW models are exact theories, and it is
remarkable that \emph{despite the absence of isometries, a similar
though weaker hierarchy holds} in that case too.

Consider the following Euclidean-signature conformal cosets
\begin{eqnarray}
    \mathrm{C}H_{d,k} &=& \frac{SO(d,1)_{-k}}{SO(d)_{-k}}\ ,\quad d=2,3,\dots
\label{ch}\\
    \mathrm{C}S_{d,k} &=& \frac{SO(d+1)_k}{SO(d)_k}\
, \quad d=2,3,\dots \label{cs1}
\end{eqnarray}
where $k$ and $-k$ indicate the level of the corresponding current
algebras and their signs ensure that the target space has
Euclidean signature. This notation reminds of the geometric cosets
but it should be kept in mind that the latter are different from
the conformal ones (gauged WZW) discussed here. Explicit forms of
the corresponding backgrounds have been worked out in the
literature for various values of $d$, both to lowest order in
$\alpha'\sim 1/k$ \cite{BCR,EFR,MSW,wittenbh,BS1,BS2,BS3} and to
all orders \cite{BSexa,DVV}.

Our observation can be formulated as follows: the radial infinity
of the space of the non-compact coset $\mathrm{C}H_{d,k+2d-4}$ is
the full space of the compact coset $\mathrm{C}S_{d-1,k}$ times a
decoupled scalar (the radial coordinate) with linear dilaton
$\mathbb{R}_{Q_{k,d}}$, where $Q_{k,d}=\frac{1-d}{2\sqrt{k+d-3}}$
is an appropriate background charge. As a consequence,
\emph{$\mathrm{C}S_{d-1,k}$ does not appear as a leaf of
$\mathrm{C}H_{d,k+2d-4}$ at finite radial coordinate but only when
this coordinate becomes infinite}. A similar property holds for
Minkowskian-signature conformal cosets $\mathrm{CAdS}_{d,k} =
\nicefrac{SO(d-1,2)_{-k}}{SO(d-1,1)_{-k}}$ or $\mathrm{CdS}_{d,k}
= \nicefrac{SO(d,1)_k}{SO(d-1,1)_k}$. In these cases the infinity
can be spatial (large radius) with time-like boundary, or it can
be temporal (remote time) with space-like boundary.

\section{An example in three dimensions}\label{sec:3and4}

The above property of conformal cosets can be proven exactly, to
all orders in $1/k$ \cite{PS06}. Here, I would like to illustrate
how the argument goes by analyzing the three-dimensional example
in the large-$k$ regime. Using global variables \cite{BS3}, the
metric and dilaton for $\mathrm{C}H_{3,k+2}$ read
\begin{eqnarray}
\mathrm{d}s_{(3)}^2&=&2 k\left({\mathrm{d}\hat{b}^2\over
4\left(\hat{b}^2-1\right)} +{\hat{b}-1\over \hat{b}+1}
{\mathrm{d}\hat{u}^2\over 4\hat{u}(\hat{v}-\hat{u}-2)}-
{\hat{b}+1\over \hat{b}-1} {\mathrm{d}\hat{v}^2\over
4\hat{v}(\hat{v}-\hat{u}-2)}\right),
\label{met3d}\\
\mathrm{e}^{-2\Phi_{(3)}} &=& \mathrm{e}^{-2\Phi_0}
\left|\left(\hat{b}^2-1\right)(\hat{v}-\hat{u}-2)\right|,\label{dil3d}
\end{eqnarray}
where $\left|\hat{b}\right|>1,0<\hat{v}<\hat{u}+2<2$. At large
$\hat{b}$ (radial coordinate), the above are better expressed in
coordinates
\begin{equation}
 \hat{b}=\exp 2x ,\quad  \hat{u}=-2 \sin^2\theta \cos^2\phi,\quad \hat{v}= 2 \sin^2\theta \sin^2\phi .
\label{fcgh1}
\end{equation}
Keeping also the subleading term (in $1/\hat{b} = \exp -2x$) in
the metric, one finds:
\begin{equation}
  \mathrm{d}s_{(3)}^2= 2k \mathrm{d}x^2 + \mathrm{d}s^2_{(2)}
+ 4k \mathrm{e}^{-2 x} \left[2 \tan\theta\, \sin 2\phi \,
\mathrm{d}\theta \, \mathrm{d}\phi -\cos2\phi
\left(\mathrm{d}\theta^2-\tan^2 \theta\,
\mathrm{d}\phi^2\right)\right] \label{cor2d}
\end{equation}
and
\begin{equation}\label{dil2d}
  \mathrm{e}^{-2\Phi_{(3)}} = 2\mathrm{e}^{4x}
  \mathrm{e}^{-2\Phi_{(2)}}  \quad \mathrm{with} \quad
  \mathrm{e}^{-2\Phi_{(2)}}=\cos^2\theta.
\end{equation}
The leading contributions
\begin{eqnarray}
  \mathrm{d}s^2_{(3-\mathrm{lead})}&=& 2k \mathrm{d}x^2 + \mathrm{d}s^2_{(2)} \nonumber \\
 &=& 2k \mathrm{d}x^2 + 2k (\mathrm{d}\theta^2 + \tan^2\theta \
  \mathrm{d}\phi^2)
\label{metdil2d}
\end{eqnarray}
and $\Phi_{(2)}$ are the background fields of
$\mathbb{R}_{Q_{3,k}}\times\mathrm{C}S_{2,k}$, as advertised
previously. The radial coordinate $x$ supports a linear dilaton
with background charge $Q_{3,k}$, which is found to be
$-1/\sqrt{k}$ by comparing the normalization of the field $x$ with
the slope of the linear dilaton (Eq. (\ref{dil2d})). With this
information, one easily checks that the central charges indeed
match, by using the general following formulas (exact in $k$):
\begin{eqnarray}
c_{{\mathrm{C}H}_{3,k+2}}&=&{6(k+2)\over k}-{3(k+2)\over
k+1},\\
c_{{\mathrm{C}S}_{2,k}}&=&{3k\over  k+1}-1,\\
 c_{\mathbb{R}_{Q_{3,k}}}&=& 1+ 12
Q_{3,k}^2= 1 + {12\over k}.
\end{eqnarray}
Details can be found in \cite{PS06}, where as already stressed,
the above arguments are shown to hold beyond the large-$k$
approximation and for all $d$.

\section{The role of parafermions}\label{sec:paraf}

At large spatial infinity, the three-dimensional exact coset
${\mathrm{C}H}_{3,k + 2}$ is factorized in an exact
two-dimensional $\mathrm{C}S_{2,k}$ times a free scalar with
background charge $\mathbb{R}_{Q_{3,k}}$. Hence, it must be
possible to dynamically generate the full three-dimensional
theory, i.e. beyond its asymptotic region, by using an integrable
marginal perturbation driven by a $(1,1)$ operator of
$\mathbb{R}_{Q_{3,k}}\times\mathrm{C}S_{2,k}$. This operator is
read off in the subleading correction of the metric (\ref{cor2d}):
\begin{equation}\label{corop}
  \delta \mathcal{L} =
   4k \mathrm{e}^{-2 x} \left[\tan\theta\, \sin 2\phi \left(
\partial_+\theta\,
\partial_-\phi + \partial_+ \phi\,
\partial_-\theta\right) -  \cos2\phi
\left(\partial_+\theta \,\partial_-\theta-\tan^2 \theta\,
\partial_+\phi \,\partial_-\phi\right)\right].
\end{equation}
The latter can indeed be reexpressed in terms of natural objects
in the $\mathbb{R}_{Q_{3,k}} \times \mathrm{C}S_{2,k}$ conformal
field theory. For the $\mathrm{C}S_{2,k}$ factor the natural
objects are the parafermions \cite{BCR}.

The semiclassical expressions for the chiral parafermions
(holomorphic) in terms of space variables are (a factor involving
$k$ is ignored)
\begin{equation}
 \Psi_\pm =
 (\partial_+ \theta \mp i \tan\theta \partial_+\phi)
%\mathrm{e}^{\mp i (\varphi -\int J^1_+ d\s^+)}\ ,
\mathrm{e}^{\mp i (\phi +\phi_1)},
 \label{ncpar}
\end{equation}
where the phase is
\begin{equation}
\phi_1  = -\frac{1}{2} \int^{\sigma^+}\! J^1_+\mathrm{d}\sigma^+
+\frac{1}{2} \int^{\sigma^-} J_-^1 \mathrm{d}\sigma^- ,\quad
J^1_\pm = \tan^2\theta \partial_\pm \phi. \label{phi}
\end{equation}
The phase obeys on-shell the condition $ \partial_+\partial_-
\phi_1 = \partial_-\partial_+ \phi_1$ and is well defined, due to
the classical equations of motion. Similarly, the expressions for
the antichiral parafermions (antiholomorphic) are
\begin{equation} \bar \Psi_\pm =
 (\partial_-\theta \pm i \tan\theta \partial_-\phi)
\mathrm{e}^{\pm i (\phi -\phi_1)} . \label{ncbpar}
\end{equation}
The exact conformal weights of the parafermions are
$\left(1-\frac{1}{2k},0\right)$ for the chiral and
$\left(0,1-\frac{1}{2k}\right)$ for the antichiral ones.

The correction (\ref{corop}) is reproduced as
\begin{equation}
\delta \mathcal{L} = -2 k  V_{3,k} \left(\Psi_+ \bar \Psi_- +
\Psi_- \bar \Psi_+\right) , \label{delScir}
\end{equation}
with $V_{3,k} = \exp-2x$ a vertex operator of weights
$(\nicefrac{1}{2k}, \nicefrac{1}{2k})$. Therefore $\delta
\mathcal{L} $ has indeed conformal weights $(1,1)$ as it should.
Its exactness is inherited from the relation established between
the two theories $\mathrm{C}H_{3,k+2}$ and $\mathrm{C}S_{2,k}
\times \mathbb{R}_{Q_{3,k}}$, which, being exact, are necessarily
connected by an integrable marginal perturbation. Notice that Eq.
(\ref{delScir}) is valid at any finite $k$ whereas
(\ref{corop})--(\ref{ncbpar}) are semiclassical expressions, as
the whole analysis of Sec. \ref{sec:3and4}.

Similar considerations hold for higher dimensions and I refer to
the already cited paper for further reading.

\section{Comments}\label{sec:co}

Ordinary FRW universes, i.e. space--times based on homogeneous
spaces, are obtained by solving string equations to some order in
$\alpha'$. Time evolution usually emerges through a time-dependent
dilaton as well as in a warping factor of the spatial metric. This
genuine time evolution is often identified at late times, with RG
evolution where the two-dimensional scale plays the role of time
\cite{ABEN,Tsey,BOP}.

In the present set-up, exact $d$-dimensional backgrounds
$\mathcal{B}$ are constructed, whose $(d-1)$-dimensional
``boundaries'' $\partial\mathcal{B}$ are also exact conformal
field theories. Supplemented with an extra free field with
background charge, the theory on $\partial\mathcal{B}$ admits a
truly marginal deformation that allows to reconstruct the theory
on $\mathcal{B}$.

If $\mathcal{B}$ is Minkowskian and if $\partial\mathcal{B}$ is
space-like, we are in a situation similar to that of the ordinary
-- inhomogeneous though -- FRW universes in the following sense:
the universe is generated by letting evolve in time some initial
space, which is per se a solution of the equations of motion in
one dimension less. This evolution is however more involved than a
simple warping (see Eq. (\ref{met3d}), which is similar to its
Minkowskian analogue -- $\mathrm{CAdS}_{d,k}$ or
$\mathrm{CdS}_{d,k}$) and \emph{can never be identified with an RG
evolution since it corresponds to a marginal deformation}. This
latter property persists of course in the case of Euclidean
$\mathcal{B}$ making it hard to speculate about holography.

Whether the above ideas could be of any practical use in string
cosmology is questionable. There is no doubt, however, that they
open yet another window to \emph{the deep problem of understanding
the very concept of time in string theory and its relation with
the Liouville field}.

The emergence of parafermions as building blocks of exactly
marginal operators, when appropriately dressed, is in the heart of
the present analysis. These appear either for the deformation of a
brane distribution or for establishing the advertised FRW-like
hierarchy in string theory. Our proof that these operators are
indeed integrable is indirect and relies on the independent
observation that they generate a line of continuous deformation
with vanishing beta-functions to all orders. A proof based on
genuine conformal-field-theory techniques would require mastering
of (non-)Abelian quantum parafermions, which is a notoriously
difficult subject.

%\vskip 0.56cm \centerline{\bf Acknowledgements} \vskip 0.25cm
%\noindent

\end{document}